\documentclass[journal,final]{IEEEtran}
\usepackage{amsmath,amsfonts}
\usepackage{amssymb}
\usepackage{amsthm}
\usepackage{algorithm}
\usepackage{array}
\usepackage[caption=false, font=footnotesize, labelfont=footnotesize, textfont=footnotesize]{subfig}
\usepackage{textcomp}
\usepackage{stfloats}
\usepackage[caption=false]{subfig}
\usepackage{url}
\usepackage{verbatim}
\usepackage{graphicx}
\usepackage{cite}
\usepackage{algorithmicx}
\usepackage{algpseudocode}
\usepackage{bm} 
\usepackage{booktabs}
\usepackage{xcolor}

\newtheorem{lemma}{Lemma}
\newtheorem{proposition}{Proposition}

\begin{document}
\bstctlcite{IEEEexample:BSTcontrol}

\title{Energy-Efficient Velocity Profile Optimization for Movable Antenna-Enabled Sensing Systems}

\author{
Jiannan~Wang, 
Yuyi~Mao,~\IEEEmembership{Senior Member,~IEEE}, 
Xianghao~Yu,~\IEEEmembership{Senior Member,~IEEE}, \\
and Ying-Jun Angela Zhang,~\IEEEmembership{Fellow,~IEEE}
    \thanks{
    
    Jiannan Wang and Xianghao Yu are with the Department of Electrical Engineering, City University of Hong Kong, Hong Kong (email: jiannan.wang@my.cityu.edu.hk, alex.yu@cityu.edu.hk). \textit{(Corresponding author: Xianghao Yu.)}
    
    Yuyi Mao is with the School of Computer Science and Engineering, Macau University of Science and Technology, Taipa, Macau, China (e-mail: yymao@must.edu.mo).

    Ying-Jun Angela Zhang is with the Department of Information Engineering, The Chinese University of Hong Kong, Hong Kong, China (e-mail: yjzhang@ie.cuhk.edu.hk).
    }
}

\maketitle

\begin{abstract}
Movable antennas (MAs) enable the reconfiguration of array geometry within a bounded region to exploit sub-wavelength spatial degrees of freedom in wireless communication and sensing systems. However, most prior research has predominantly focused on the communication and sensing performance, overlooking the mechanical power consumption inherent in antenna movement. To bridge this gap, this paper investigates a velocity profile optimization framework for MA-assisted direction-of-arrival (DoA) estimation, explicitly balancing sensing accuracy with mechanical energy consumption of MAs. We first establish a Newtonian-based mechanical energy model, and formulate a functional optimization problem for sensing energy efficiency (EE) maximization. By applying the calculus of variations, this formulation is transformed into an infinite-dimensional problem defined by the Euler-Lagrange equation. To solve it, we propose a spectral discretization framework based on the Galerkin method, which expands the velocity profile over a sinusoidal basis. In the regime where energy consumption is dominated by linear damping, we prove that the optimal velocity profile follows a closed-form sinusoidal shape. For more general scenarios involving strong nonlinear aerodynamic drag, we leverage the Markov-Lukács theorem to transform the kinematic constraints into strictly convex sum-of-squares (SOS) conditions. Consequently, the infinite-dimensional problem is reformulated as a tractable finite-dimensional nonlinear algebraic system, which is solved by a two-layer algorithm combining Dinkelbach’s method with successive convex approximation (SCA). Numerical results demonstrate that our optimized velocity profile significantly outperforms baselines in terms of EE across various system configurations. Insights into the optimized velocity profiles and practical design guidelines are also provided.
\end{abstract}

\begin{IEEEkeywords}
Energy efficiency, functional optimization, movable antenna, spectral discretization.
\end{IEEEkeywords}

\section{Introduction}
\bstctlcite{IEEEexample:BSTcontrol}
\label{sec:introduction}
The transition toward sixth-generation (6G) wireless networks is driven by the need to extend conventional fifth-generation (5G) services into further enhanced mobile broadband (eMBB), ultra massive machine-type communications (umMTC), and enhanced ultra-reliable and low-latency communications (eURLLC)~\cite{10379539}. These advanced capabilities constitute the cornerstone for realizing transformative applications, including high-precision environmental sensing, immersive holographic communications, and large-scale digital twins~\cite{10054381}. To satisfy these emerging demands, one of the key challenges lies in efficiently exploiting the spatial degrees of freedom (DoFs) inherent in the wireless channel. To this end, significant research efforts have been dedicated to expanding the spatial dimensions of communication systems through technologies such as extremely large-scale multiple-input multiple-output (XL-MIMO)~\cite{7397861,7400949} and reconfigurable intelligent surfaces (RIS)~\cite{8811733,9140329,9206044}. By deploying massive antenna arrays or programmable reflective elements, these paradigms enhance coverage and combat path loss by leveraging high beamforming gains and establishing virtual line-of-sight (LoS) links, respectively. However, these architectures are fundamentally limited by the fixed-position antenna (FPA) configuration. This static deployment prevents transceivers from adapting to fine-grained small-scale fading variations, thereby restricting wireless networks from fully exploiting the spatial diversity available in complex scattering environments.

To address the limitations of FPAs, a set of closely related technologies, including movable antennas (MAs)~\cite{10906511}, fluid antenna systems (FAS)~\cite{10767351}, pinching antennas~\cite{11111701}, and moving arrays~\cite{10777052}, have emerged as promising solutions to unlock additional spatial DoFs. Unlike traditional FPAs, these novel antenna configurations employ reconfigurable mechanisms that allow antennas to adjust their positions within a bounded region in order to adapt to dynamic wireless channels. The authors of~\cite{10318061} revealed that even sub-wavelength displacements enable MAs to effectively harness small-scale fading, yielding substantial performance gains over FPAs. Motivated by these intrinsic advantages, extensive research efforts have been recently devoted to leveraging MAs for enhancing the performance of various wireless communication and sensing systems. For example, in terms of wireless communication, \cite{10694747} and \cite{10741192} have investigated the joint design of MA positions and beamforming strategies to maximize the achievable rates in the multicast and uplink transmission scenarios, respectively. In addition, for wireless sensing or integrated sensing and communication (ISAC) systems, there have been many research efforts~\cite{11086422,10696953,10737418} on designing the array geometry to optimize the Cramér-Rao bound (CRB) or the trade-off between sensing accuracy and communication rate.

Despite the significant performance gains achieved by MAs, their practical deployment is constrained by the energy overhead associated with mechanical reconfiguration. Consequently, balancing the trade-off between system performance and power consumption has become a critical design objective. For communication-centric systems, energy efficiency (EE) is typically formulated as the ratio of the achievable sum-rate to the consumed power~\cite{10845796,10681491,11216118,11130638,11321281,11223265}. For sensing-centric systems, the focus has shifted towards defining EE metrics based on estimation accuracy, utilizing objective functions such as the ratio of Fisher information or sensing mutual information to the total power consumption~\cite{10445319,10965357,11168825}. While these metrics provide a standardized evaluation framework, existing research generally exhibits limitations in two critical aspects, i.e., the neglect or oversimplification of mechanical power modeling and the lack of velocity profile optimization.

First, regarding energy consumption modeling, most existing works rely on static or simplified movement-based approximations that fail to capture the kinetics of antenna movement. For instance, \cite{11216118} modeled the MA power consumption as a constant value proportional to the number of active antennas, treating it as independent of motion states. The authors of~\cite{11130638} addressed user fairness by adopting a stepper motor model, but they assumed the movement energy consumption is proportional to the displacement, ignoring the impact of velocity variations. Similarly, the achievable sum secrecy EE, i.e., the ratio of the achievable sum secrecy rate to the energy consumption, was maximized in \cite{11321281} by combining an aerodynamic model for unmanned aerial vehicle (UAV) propulsion with a power model that assumes linear dependence on the angular displacement of MAs. Although~\cite{11223265} modeled movement energy consumption as a linear function of antenna displacement, it considered the velocity of antenna movement as a fixed parameter rather than a controllable variable. Parallel limitations persist in the research of sensing-oriented MA-enabled wireless systems. Related works such as~\cite{10445319} and \cite{10965357} optimized sensing EE while neglecting the mechanical energy dissipation from antenna movement, which concentrate exclusively on transmit power and circuit power. Likewise, the authors of~\cite{11168825} optimized sensing CRBs via power splitting, yet their energy model only considers transmit and circuit power, neglecting the mechanical cost of antenna movement.

Second, in terms of movement dynamics, existing research has largely overlooked the fine-grained design of time-varying velocity profiles under EE-oriented objectives. On the one hand, several studies have optimized the movement strategies to enhance system performance. For instance, \cite{ma2025movable} adopted a slot-based motion model where the MA velocity is optimized to minimize the CRB for angle estimation. Similarly, \cite{liu2025near} utilized functional analysis to characterize the antenna movement, deriving a closed-form optimal placement for near-field communications. However, both studies are performance-oriented and disregard the mechanical energy required to support such velocity variations. On the other hand, respecting the energy constraints, recent works have begun to address the coupling between velocity and power consumption of antenna movements. For example, although~\cite{11048972} incorporated both the MA driver’s energy expenditure and data transmission energy into its total energy consumption model, it assumed a constant and pre-defined moving velocity of MA. Similarly, \cite{wei2025energy} formulated an energy consumption model in terms of the constant MA velocity. However, this approach simplified velocity to a \textit{scalar} variable, restricting the MA system to uniform motion. Consequently, current approaches fail to fully leverage the fine-grained DoFs offered by \emph{time-varying} velocity design, which is essential for exploiting the relationship between motion and energy consumption.

In order to design energy-efficient MA systems and gain critical insights on the system design, it is imperative to shift the design perspective from static placement and simplistic motion models to dynamic velocity profile optimization. To bridge this gap, this paper formulates a velocity profile optimization framework for MA-enabled sensing systems, where the consumed energy is modeled as a function of the velocity of the MA. The main contributions of this paper are summarized in four aspects as follows:
\begin{itemize}
    \item \textit{System modeling and formulation}: We establish a signal model for MA-based one-dimensional (1-D) direction-of-arrival (DoA) estimation that explicitly accounts for the antenna's time-varying velocity. In addition, based on Newtonian mechanics, we introduce a mechanical energy model for MAs, incorporating inertial forces, linear damping, and quadratic aerodynamic drag induced by antenna movement. Then, a functional optimization problem is formulated to maximize the sensing EE subject to kinematic constraints on velocity and track length.
    
    \item \textit{Spectral discretization and analytical characterization}: To tackle the infinite-dimensional nature of the formulated problem, we propose a spectral discretization framework based on the Galerkin method. Specifically, we expand the velocity profile over a sinusoidal basis, transforming the variational problem into a finite-dimensional algebraic system. Based on this framework, for the regime where energy consumption is dominated by linear damping, we theoretically prove that the optimal velocity profile follows a closed-form sinusoidal solution.

    \item \textit{Two-layer algorithm for numerical optimization}: For the general scenario involving strong nonlinear aerodynamic drag, where analytical solutions are intractable, we develop a numerical optimization algorithm. We utilize the Markov-Lukács Theorem to transform the kinematic constraints into strictly convex sum-of-squares (SOS) conditions in terms of coefficients. To solve the formulated problem, we develop a two-layer algorithm combining Dinkelbach’s method and successive convex approximation (SCA), which effectively transforms the intractable non-convex fractional objective into a sequence of tractable convex subproblems.

    \item \textit{Numerical verification and insights}: Simulations demonstrate the advantages of the proposed method over conventional uniform and trapezoidal motion benchmarks. By fully exploiting the DoFs in velocity shaping, our approach adaptively mitigates the impact of nonlinear aerodynamic drag, yielding a superior sensing-energy trade-off. Notably, in the linear damping-dominated regime, the numerical solution converges to the derived analytical solution, eliminating the need for iterative computation. Furthermore, we provide some guidelines for MA deployments, demonstrating that energy-efficient sensing necessitates delicately designed velocity profiles and a balance between the track length and the sensing interval.
\end{itemize}


\textit{Notations}: Scalars, vectors, and matrices are denoted by lower-case $x$, bold-case lower-case $\mathbf{x}$, and bold-case upper-case letters $\mathbf{X}$, respectively. The transpose of a matrix $\mathbf{X}$ is denoted by $\mathbf{X}^\mathrm{T}$. The operator $\mathrm{diag}(\cdot)$ constructs a diagonal matrix with the arguments as its diagonal entries, and $\mathbf{I}_n$ denotes the $n$-th order identity matrix. $\mathbb{R}^{M \times N}$ and $\mathbb{C}^{M \times N}$ denote the spaces of $M \times N$ real and complex matrices, respectively. $\mathbb{S}^N$ represents the set of symmetric $N \times N$ matrices, and $\mathbf{X} \succeq 0$ indicates that $\mathbf{X}$ is positive semidefinite (PSD). The Euclidean norm of a vector $\mathbf{x}$ is denoted by $\|\mathbf{x}\|_2$. $\mathbb{E}\{\cdot\}$ denotes the statistical expectation operator. For a continuous function $f(t)$, $\dot{f}(t)$ denotes its first-order derivative. In addition, $\binom{n}{k} = \frac{n!}{k!(n-k)!}$ denotes the binomial coefficient, $\deg(\cdot)$ returns the degree of a polynomial, and $\lfloor \cdot \rfloor$ is the floor function.

\section{System Model} 
\label{sec:system_model} 
This section establishes the signal model for the MA-enabled sensing system and derives the performance metric for DoA estimation.
\subsection{Signal Model}
\label{sec:signal_model}
\begin{figure}[!t]
    \centering    \includegraphics[width=0.45\textwidth]{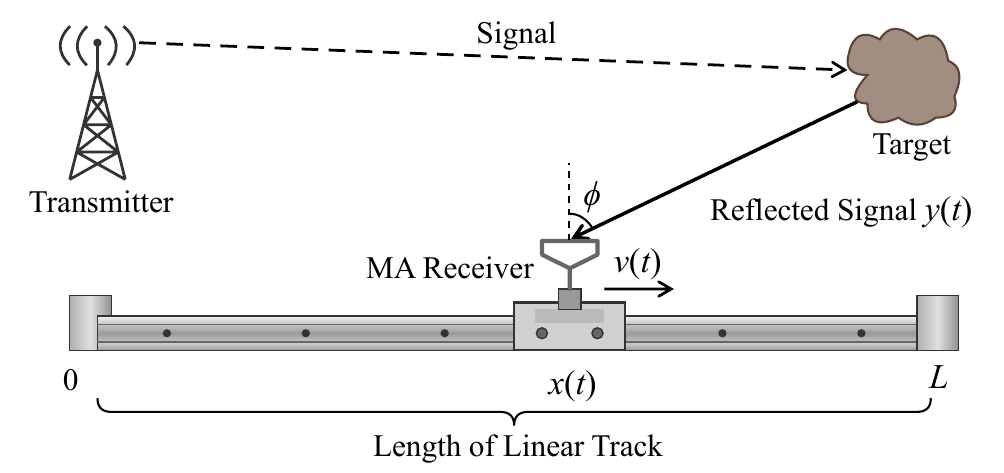} 
    \caption{Diagram of a bi-static wireless sensing system with a single MA.}
    \label{fig:system_model}
\end{figure}

Consider a single MA moving along a one-dimensional linear track of length $L$ over a finite sensing interval $T$, as illustrated in Fig. \ref{fig:system_model}. Let $x(t)$ and $v(t)$ denote the instantaneous position and velocity of the antenna at time $t$, respectively. Without loss of generality, the antenna is assumed to move from the origin with zero initial velocity, i.e., $x(0)=0$ and $v(0)=0$. Consequently, the trajectory $x(t)$ is determined by the integral of its velocity profile $v(t)$ as
\begin{equation}
\label{eq: 1}
    x(t) = \int_{0}^{t} v(\tau) \, \mathrm{d}\tau, \quad t \in [0, T].
\end{equation}

We consider a bistatic sensing for DoA estimation, where the transmitter is spatially separated from the MA receiver. The transmitter illuminates a target, which subsequently reflects the signal to the MA receiver. Then, the continuous-time received signal $y(t)$ can be modeled as
\begin{equation}
\label{eq: c_signal_model}
    y(t) = \alpha  e^{-\jmath \frac{2\pi}{\lambda} x(t) \theta} s + n(t),
\end{equation}
where $\theta = \cos(\phi) \in [-1, 1]$ denotes the directional cosine of the unknown DoA with physical angle $\phi$. Furthermore, $\alpha \in \mathbb{C}$ represents the complex channel gain, $s$ denotes the known pilot signal with power $P_s$, and $\lambda$ is the carrier wavelength. The term $n(t) \sim \mathcal{CN}(0, \sigma_{\mathrm{n}}^2)$ is the complex-valued additive white Gaussian noise (AWGN) with power $\sigma_{\mathrm{n}}^2$. 

To facilitate the digital signal processing, the continuous-time signal $y(t)$ is uniformly sampled at $M$ discrete time instants $\mathbf{t}=[t_1, t_2, \cdots, t_M]^\mathrm{T}$ where $t_m=\frac{Tm}{M}$ with $m=1,2,\cdots,M$. Then, the resulting discrete received signal vector $\mathbf{y}\in \mathbb{C}^{M \times 1}$ is given by
\begin{equation}
\label{eq: d_signal_model}
    \mathbf{y} = \alpha \mathbf{b}(\theta) s + \mathbf{n},
\end{equation}
where $\mathbf{n} \sim \mathcal{CN}(0, \sigma_\mathrm{n}^2\mathbf{I}_M)$ is the AWGN vector. The term $\mathbf{b}(\theta) \in \mathbb{C}^{M\times1}$ denotes the steering vector associated with the antenna positions, which is given by
\begin{equation}
\label{eq: steering_vector}
\mathbf{b}(\theta) = \left[e^{-\jmath \frac{2\pi}{\lambda} x(t_1) \theta}, e^{-\jmath \frac{2\pi}{\lambda} x(t_2) \theta}, \cdots, e^{-\jmath \frac{2\pi}{\lambda} x(t_M) \theta}\right]^\mathrm{T}.
\end{equation}
Here, $\mathbf{x}=[x(t_1), x(t_2), \cdots, x(t_M)]^\mathrm{T}$ denotes the discrete antenna position vector (APV) with respect to $M$ snapshots. With the signal model established in \eqref{eq: d_signal_model}, one can apply the maximum likelihood \cite{57542} or fast Fourier transform \cite{7815358} to estimate the DoA. However, it is unclear how the antenna's movement strategy affects the estimation performance of DoA. To answer this, we analyze the CRB in the next subsection.

\subsection{CRB of 1-D DoA Estimation}
\label{sec: crb}
The estimation accuracy of parameter $\theta$ is theoretically lower-bounded by the CRB, which is defined as the inverse of the Fisher information. For any unbiased estimator $\hat{\theta}$, the mean square error (MSE) satisfies the following inequality~\cite{ma2025movable,poor2013intro}
\begin{equation}
\label{eq: crb}
    \mathbb{E}\{(\hat{\theta}-\theta)^2\} \geq \text{CRB}(\theta) = \frac{\sigma_{\mathrm{n}}^2 \lambda^2}{8\pi^2TP_sM|\alpha|^2} \frac{1}{\text{var}(\mathbf{x})},
\end{equation}
where $\text{var}(\mathbf{x})$ denotes the sample variance of the elements in APV, given by
\begin{equation}
    \mathrm{var}(\mathbf{x}) = \frac{1}{M}\sum_{m=1}^{M}x^2(t_m)-
    \left(
    \frac{1}{M}\sum_{m=1}^{M}x(t_m)
    \right)^2.
\end{equation}

The formulation above reveals that minimizing the CRB is equivalent to maximizing the variance $\text{var}(\mathbf{x})$. Therefore, one could optimize the discrete APV $\mathbf{x}$ by driving the antenna to concentrate at the track boundaries to maximize the spatial spread. However, physically realizing such a distribution would require the MA to perform rapid transitions between endpoints, which inevitably demands an exceedingly high velocity, especially in the middle region of the track.

To guarantee kinematic feasibility and fully leverage the DoFs offered by the design of the velocity profile, we shift the optimization variable from discrete positions $\mathbf{x}$ to the time-varying velocity profile $v(t)$. Notably, as the number of snapshots $M$ increases, $\mathrm{var}(\mathbf{x})$ can be asymptotically approximated by the following continuous variance functional
\begin{equation}
\label{eq: var}
    \mathcal{V}(v)
    =
    \frac{1}{T}\int_{0}^{T}x^2(t)\,\mathrm{d}t
    -
    \left(
    \frac{1}{T}\int_{0}^{T}x(t)\,\mathrm{d}t
    \right)^2.
\end{equation}

According to \eqref{eq: 1} and \eqref{eq: var}, scaling the velocity profile as $v_{\rho}(t)=\rho v(t)$ with $\rho>0$ leads to $\mathcal{V}(v_{\rho})=\rho^2\mathcal{V}(v)$. Therefore, for a given velocity profile shape, increasing the velocity magnitude enlarges the spatial variance. However, maintaining such high speeds incurs prohibitive mechanical energy costs. This inherent trade-off between the sensing accuracy and energy consumption motivates the study of energy-efficient MA systems, detailed in the following section.

\section{Problem Formulation}
\label{sec:problem_formulation}
While maximizing the position variance $\mathcal{V}(v)$ enhances the sensing performance, it also imposes significant mechanical energy costs. In this section, we give the energy consumption model and formulate the velocity profile optimization problem.

\subsection{Mechanical Energy Consumption Model}
From the perspective of Newtonian mechanics, an antenna of mass $m_{\mathrm{a}}$ moving along a linear track requires an actuator force $F(t)$ to overcome inertial forces, viscous friction, and aerodynamic drag~\cite{8663615,mccormick1994aerodynamics}. The dynamic force equation is modeled as
\begin{equation}
    F(t) = m_{\mathrm{a}} \dot{v}(t) + \alpha_1 v(t) + \alpha_2 v^2(t),
\end{equation}
where $\dot{v}(t)$ denotes the acceleration, while $\alpha_1$ and $\alpha_2$ represent the linear damping and quadratic drag coefficients, respectively. The first term, $m_{\mathrm{a}} \dot{v}(t)$, represents the inertial force required to change the antenna’s momentum. The second term, $\alpha_1 v(t)$, models a linear viscous friction (e.g., from mechanical bearings or internal damping) that is proportional to velocity. The third term, $\alpha_2 v^2(t)$, corresponds to quadratic aerodynamic drag, which dominates at high speeds and depends on the frontal area, air density, and drag coefficient of the antenna assembly. In addition, we consider a velocity-independent static power consumption $P_{\mathrm{s}}$ that accounts for the baseline energy expenditure of the sensing system during operation~\cite{10445319,10965357,11168825}. Then, the instantaneous mechanical power of the system is  $P(t) = F(t)v(t)+P_{\mathrm{s}}$. Accordingly, the total energy consumption functional $\mathcal{E}(v)$ over the observation interval $[0, T]$ is derived as\footnote{The model in \eqref{eq: energy_consumption} characterizes the net mechanical work, while energy conversion losses at the actuator side are omitted in this work.}
\begin{equation}
\label{eq: energy_consumption}
\begin{aligned}
    \mathcal{E}(v) &= \int_{0}^{T} \left( P_{\mathrm{s}} + m_{\mathrm{a}} v(t)\dot{v}(t) + \alpha_1 v^2(t) + \alpha_2 v^3(t) \right) \,\mathrm{d}t \\
    &= E_{\mathrm{s}} + \frac{1}{2}m_{\mathrm{a}} v^2(T) + \int_{0}^{T} \left( \alpha_1 v^2(t) + \alpha_2 v^3(t) \right) \,\mathrm{d}t,
\end{aligned}
\end{equation}
where the equality follows from the initial condition $v(0)=0$. Besides, $E_{\mathrm{s}} \triangleq P_{\mathrm{s}} T$ denotes the static energy consumption. It can be observed from (9) that energy minimization favors low velocities. Therefore, reducing the antenna velocity decreases the energy consumption but simultaneously degrades the sensing performance, revealing a fundamental trade-off between sensing accuracy and energy consumption.

\subsection{Optimization Problem Statement}
\label{sec: dinklebach}
To quantify the trade-off between the estimation accuracy and mechanical energy consumption, we utilize the sensing EE metric. Defined as the ratio of the Fisher information to the mechanical energy cost, sensing EE characterizes the amount of spatial information acquired per unit of energy expenditure, given by~\cite{10445319,10965357,11168825}
\begin{equation}
    \text{EE}_{\mathrm{s}} = \frac{\text{CRB}^{-1}(\theta)}{\mathcal{E}(v)},
\end{equation}
whose unit is $1/\text{rad}^2/\text{Joule}$. Although maximizing the sensing EE balances sensing accuracy and energy consumption, it does not explicitly guarantee a minimum sensing quality. Therefore, we impose a sensing quality of service (QoS) requirement by constraining the spatial variance to be no smaller than a fraction $\eta$ of its theoretical upper bound $L^2/4$, where $L^2/4$ follows from Popoviciu's inequality~\cite{bhatia2000better} and $\eta\in(0,1]$ is a predefined scaling constant. Incorporating this sensing QoS requirement, together with the kinematic and travel distance constraints of the MA, the velocity profile optimization problem is formulated as
\begin{subequations}
\label{eq:opt_problem}
\begin{align}
    \max_{v(t)} \quad & \tilde{\mathcal{J}}(v) =
    \frac{\mathcal{V}(v)}{\mathcal{E}(v)} \\
    \text{s.t.} \quad & v(0) = 0, \label{eq:con_init} \\
    & 0 \le v(t) \le V_{\max}, \quad \forall t \in [0, T], \label{eq:con_ineq} \\
    & \int_{0}^{T} v(\tau) \, \mathrm{d}\tau \le L, \label{eq:con_dist} \\
    & \mathcal{V}(v) \ge \eta \frac{L^2}{4}, \label{eq:QoS}
\end{align}
\end{subequations}
where $V_{\max}$ denotes the maximum speed and the constant scaling factors in \eqref{eq: crb} are omitted in the objective function without loss of generality. The unidirectional constraint $v(t) \ge 0$ is because the sensing aperture is strictly dictated by the maximum displacement. Any direction reversal merely re-samples traversed coordinates, which yields no improvement in spatial variance but increases the energy consumption. Notably, the formulated problem presents several mathematical challenges. First, unlike conventional vector optimization where decision variables are finite-dimensional vectors, problem in \eqref{eq:opt_problem} is a functional optimization problem~\cite{liu2025near,ortega1998euler}, whose decision variable $v(t)$ resides in an infinite-dimensional function space. Consequently, standard gradient-based methods defined on Euclidean spaces are inapplicable. Second, the objective function exhibits a non-convex fractional structure, coupled with highly nonlinear dependencies of $\mathcal{V}(v)$ on the integral of $v(t)$. 

To tackle the fractional nature of the objective, we begin by employing Dinkelbach’s algorithm~\cite{schaible1976fractional}. This approach transforms the original fractional problem into a sequence of parametric subtractive subproblems. Specifically, we consider the following auxiliary problem
\begin{subequations}
\label{eq:new_opt_problem}
\begin{align}
    \max_{v(t), \xi} \quad & \mathcal{J}(v) =
    \mathcal{V}(v)-\xi\mathcal{E}(v) \\
    \text{s.t.} \quad & \eqref{eq:con_init}-\eqref{eq:QoS}
\end{align}
\end{subequations}
where $\xi>0$ is an auxiliary variable that balances the priority between variance maximization and energy conservation. The optimal solution to the original fractional problem in \eqref{eq:opt_problem} is recovered when the auxiliary objective function $\mathcal{J}(v)$ vanishes, at which point $\xi$ equals the maximal EE. Thus, the optimization task simplifies to iteratively solving the subproblem in \eqref{eq:new_opt_problem}. Since this subproblem remains a functional optimization task in an infinite-dimensional space, standard vector optimization frameworks are inapplicable. In the following section, we apply the calculus of variations~\cite{ortega1998euler} to tackle the problem in \eqref{eq:new_opt_problem}.

\section{Functional Analysis and Discretization}
\label{sec:functional_analysis}
This section presents a functional analysis of the optimization problem in \eqref{eq:new_opt_problem}. We first derive the Euler-Lagrange equation. Then, we reformulate the equation into a tractable finite-dimensional algebraic system using the Galerkin-based weighted residual method (WRM).
\subsection{Variational Derivative and Euler-Lagrange Equation}
\label{sec:Euler}
The derivation of the optimality condition begins with computing the variational derivative. Prior to deriving the variation of the objective function, it should be noted that the variance functional $\mathcal{V}(v)$ defined in \eqref{eq: var} is not an explicit function of $v(t)$. To streamline the derivation, we first present the following proposition.
\begin{proposition}
\label{proposition1}
The variance functional $\mathcal{V}(v)$ defined in \eqref{eq: var} can be reformulated as an explicit kernel representation in terms of $v(t)$, which is given by
\begin{equation}
\label{eq: kernel_rep_v}
    \mathcal{V}(v) = \iint\limits_{[0,T]^2} v(u) K(u,s) v(s)  \, \mathrm{d}u \, \mathrm{d}s,
\end{equation}
where the kernel function $K(u,s)$ is defined as
\begin{equation}
\label{eq:kernel}
    K(u,s) = \frac{T \min(u,s) - us}{T^2}.
\end{equation}

Proof: Please see Appendix \ref{sec:app1}.
\end{proposition}

From Proposition \ref{proposition1}, the variance functional $\mathcal{V}(v)$ is a quadratic form of $v(t)$ induced by the symmetric kernel $K(u,s)$. Notably, the kernel $K(u,s)$ is proportional to the covariance function of a standard Brownian bridge process \cite{van2004detection}. This structure is pivotal for the spectral analysis in the next subsection. Leveraging the simplified and explicit expression of the variance functional in \eqref{eq:kernel}, we can efficiently seek the stationary point of $\mathcal{J}(v)$ at which its first-order variation vanishes. According to the definition of the variational derivative, we introduce a perturbed velocity function as
\begin{equation}
    \tilde{v}(t) = v(t) + \epsilon \eta(t),
\end{equation}
where $\epsilon$ is a scalar representing the magnitude of the perturbation, and $\eta(t)$ is an arbitrary differentiable function. Since the velocity is fixed at the initial time $t=0$ but free at the terminal time $t=T$, the admissible variation $\eta(t)$ is required to satisfy $\eta(0)=0$ to preserve the initial condition $v(0)=0$, while $\eta(T)$ remains unconstrained. We first derive the variational derivative of $\mathcal{V}(v)$. Substituting the perturbed velocity into the kernel representation in \eqref{eq: kernel_rep_v}, one has
\begin{equation}
\begin{aligned}
    \mathcal{V}(\tilde{v}) 
    &= \iint\limits_{[0,T]^2} (v(u)+\epsilon\eta(u))K(u,s)(v(s)+\epsilon\eta(s)) \, \mathrm{d}u \, \mathrm{d}s  \\
    &= \mathcal{V}(v) + 2\epsilon \iint\limits_{[0,T]^2} v(u)K(u,s)\eta(s) \, \mathrm{d}u \, \mathrm{d}s + \mathcal{O}(\epsilon^2),
\end{aligned}
\end{equation}
where the symmetry of the kernel, i.e., $K(u,s) = K(s,u)$, is exploited to combine the linear terms. In addition, $\mathcal{O}(\epsilon^2)$ represents higher-order terms involving $\epsilon^2$. Then, the first variation of $\mathcal{V}(v)$ with respect to $v(t)$ can be obtained as
\begin{equation}
\label{eq: grad1}
    \delta \mathcal{V} = \left. \frac{\mathrm{d} \mathcal{V}(\tilde{v})}{\mathrm{d}\epsilon} \right|_{\epsilon=0} 
    = 2 \iint\limits_{[0,T]^2} v(u)K(u,s)\eta(s) \, \mathrm{d}u \, \mathrm{d}s.
\end{equation}

Next, we derive the variation of the energy consumption term $\mathcal{E}(v)$ defined in \eqref{eq: energy_consumption}. Following a similar procedure, we obtain the variation $\delta \mathcal{E}$ as
\begin{equation}
\begin{aligned}
\label{eq:grad2}
    \delta \mathcal{E} 
    &= \frac{\mathrm{d}}{\mathrm{d}\epsilon} \left. \left[ 
        \begin{aligned}
            & E_{\mathrm{s}} + \frac{m_{\mathrm{a}}}{2}\bigl(v(T)+\epsilon\eta(T)\bigr)^2 \\
            & + \int_0^T\Bigl(\alpha_1(v+\epsilon\eta)^2+\alpha_2(v+\epsilon\eta)^3\Bigr)\,\mathrm{d}t
        \end{aligned}
    \right] \right|_{\epsilon=0} \\
    &= m_{\mathrm{a}}v(T)\eta(T)+\int_0^T\Bigl(2\alpha_1v(t)+3\alpha_2v^2(t)\Bigr)\eta(t)\,\mathrm{d}t.
\end{aligned}
\end{equation}

Finally, combining the results in \eqref{eq: grad1} and \eqref{eq:grad2}, the first variation of the total objective function $\mathcal{J}(v) = \mathcal{V}(v) - \xi \mathcal{E}(v)$ can be given by
\begin{equation}
\label{eq: total_variation}
    \delta \mathcal{J} = \int_0^T \Phi(t) \eta(t)\,\mathrm{d}t - \xi m_{\mathrm{a}} v(T) \eta(T),
\end{equation}
where the integrand $\Phi(t)$ is given by
\begin{equation}
    \Phi(t) = 2 \int_0^T K(t,s)v(s)\,\mathrm{d}s - \xi(2\alpha_1 v(t) + 3\alpha_2 v^2(t)).
\end{equation}

For $v(t)$ to be a stationary point, $\delta \mathcal{J}$ must vanish for every admissible perturbation $\eta(t)$. This yields two sets of conditions. First, since $\eta(t)$ is arbitrary within the interval $[0,T]$, the integrand $\Phi(t)$ must vanish pointwise, yielding the well-known Euler-Lagrange equation \cite{ortega1998euler} as
\begin{equation}
\label{eq:EL_equation}
    2 \int_{0}^{T} K(t, s) v(s) \,\mathrm{d}s = \xi \left( 2\alpha_1 v(t) + 3\alpha_2 v^2(t) \right).
\end{equation}

Second, since the perturbation at the endpoint $\eta(T)$ is arbitrary, the boundary term must vanish independently, which requires $v(T)=0$. Thus, the optimality condition implies that the antenna must come to a stop at the end of the observation period. Consequently, the optimal velocity profile is obtained by solving the nonlinear integral equation \eqref{eq:EL_equation} subject to the boundary conditions $v(0)=v(T)=0$. The algorithm to solve this integral equation is presented in the following subsection.

\subsection{Galerkin-Based Spectral Discretization}
\label{sec:spectral_discretization}
The Euler-Lagrange equation derived in \eqref{eq:EL_equation} is the Fredholm integral equation of the second kind~\cite{atkinson2009numerical}. However, the presence of the quadratic velocity term $v^2(t)$ renders a closed-form analytical solution intractable. To address this challenge, we resort to the WRM to obtain an approximate analytical solution~\cite{atkinson2009numerical,Spectral988}, which typically proceeds in three steps. First, a trial solution is constructed that satisfies the boundary conditions but contains unknown coefficients. Then, the trial solution is substituted back into the original integral equation, which results in a non-zero residual because the solution is not exact. Finally, the unknown coefficients are determined by forcing a weighted integral of this residual over the entire domain to be zero.

Based on the above introduction, the computational efficiency of the WRM critically depends on the choice of the trial function basis. To avoid computationally expensive numerical integrations involved in the term $\int_0^T K(t,s) v(s) \,\mathrm{d}s$, we seek a basis that can diagonalize this integral operator. To this end, we invoke the spectral decomposition of the kernel $K(t,s)$. As defined in \eqref{eq:kernel}, $K(t,s)$ is symmetric, continuous, and positive definite, which admits an eigendecomposition that decouples the integral operation according to the following lemma.

\begin{lemma}[Mercer Series of the Variance Kernel \cite{van2004detection}]
\label{lemma:mercer}
    Let $K(t,s)$ be the Brownian bridge kernel defined on $[0,T]^2$, then $K(t,s)$ admits the following absolutely and uniformly convergent expansion
    \begin{equation}
    \label{eq:mercer-expansion}
    K(t, s) = \sum_{n=1}^{\infty} \lambda_n \, \phi_n(t) \, \phi_n(s),
    \end{equation}
    where $\{\lambda_n\}_{n=1}^\infty$ and $\{\phi_n\}_{n=1}^\infty$ are the eigenvalues and eigenfunctions (also referred to as spectral modes) of the corresponding integral operator, characterized by the relation $\int_0^T K(t, s) \phi_n(s) \, \mathrm{d}s = \lambda_n \phi_n(t)$. The eigenvalues and eigenfunctions are given in the closed-form expression as
    \begin{equation}
    \label{eq:eigenvalues-eigenfunctions}
    \lambda_n = \frac{T}{(n\pi)^2}, \quad 
    \phi_n(t) = \sqrt{\frac{2}{T}} \, \sin\!\left( \frac{n \pi t}{T} \right),
    \end{equation}
    respectively. Here, the sine basis $\{\phi_n\}_{n=1}^\infty$ forms a complete orthogonal set in $L^2[0,T]$, satisfying $\int_0^T \phi_n(t) \phi_m(t) \, \mathrm{d}t = 1$ if and only if $n=m$ and otherwise $\int_0^T \phi_n(t) \phi_m(t) \, \mathrm{d}t = 0$.
\end{lemma}

Based on the spectral decomposition established in Lemma \ref{lemma:mercer}, we propose to approximate the optimal velocity profile $v(t)$ using a truncated series of the first $N$ spectral modes. The trial solution is constructed as
\begin{equation}
\label{eq:v_approx}
    v_N(t) = \sum_{n=1}^{N} c_n \phi_n(t),
\end{equation}
where $\mathbf{c} = [c_1, c_2, \dots, c_N]^\mathrm{T}$ are the unknown spectral coefficients to be determined. The selection of this trial function is strategic for two key reasons. First, these eigenfunctions diagonalize the integral operator, which significantly simplifies the computational complexity. Second, and critically, the sinusoidal eigenfunctions defined in \eqref{eq:eigenvalues-eigenfunctions} naturally satisfy the zero-velocity boundary conditions, i.e., $\phi_n(0)=\phi_n(T)=0$ for all $n$. Consequently, the constructed trial solution $v_N(t)$ is inherently subject to the physical constraints $v(0)=v(T)=0$ without requiring additional Lagrange multipliers.

By leveraging the trial solution in \eqref{eq:v_approx} and exploiting the orthogonality of the eigenfunctions, the left-hand side of the Euler-Lagrange equation in \eqref{eq:EL_equation} is recast as
\begin{equation}
\begin{aligned}
\label{eq: 25}
    2 \int_0^T K(t,s) v_N(s) \,\mathrm{d}s &= 
    2  \sum_{n=1}^{N} c_n\int_0^T K(t,s) \phi_n(s) \,\mathrm{d}s \\
    &= 2 \sum_{n=1}^{N} c_n \lambda_n \phi_n(t),
\end{aligned}
\end{equation}
based on which we define the residual term as the difference between \eqref{eq: 25} and the right-hand side of \eqref{eq:EL_equation}, i.e.,
\begin{equation}
\begin{aligned}
    R(t;\mathbf{c}) &= 2 \sum_{n=1}^{N} \lambda_n c_n \phi_n(t) - 2 \xi \alpha_1 \sum_{n=1}^{N} c_n \phi_n(t) \\& \quad - 3 \xi \alpha_2 \left( \sum_{n=1}^{N} c_n \phi_n(t) \right)^2.
\end{aligned}
\end{equation}

Ideally, if $v_N(t)$ were the exact solution, the residual $R(t;\mathbf{c})$ would vanish identically over the entire domain. However, due to the series truncation, a non-zero residual is inevitable. To determine the optimal coefficients $\mathbf{c}$, we employ the Galerkin method \cite{Spectral988}. The rationale for this choice lies in approximation theory. By enforcing the residual to be orthogonal to the subspace spanned by the basis functions, the Galerkin method ensures that the approximation error is minimized in the $\ell_2$-norm, yielding the optimal solution within the finite-dimensional subspace defined by the truncated series.

\begin{figure*}[!t]
\centering
\begin{equation}
\label{eq:algebraic_system}
\begin{aligned}
\int_0^T R(t;\mathbf{c}) \phi_k(t) \,\mathrm{d}t 
&= 2 \sum_{n=1}^{N} (\lambda_n - \xi \alpha_1) c_n \int_0^T \phi_n(t) \phi_k(t) \,\mathrm{d}t 
- 3 \xi \alpha_2 \sum_{n=1}^N \sum_{m=1}^N c_n c_m \int_0^T \phi_n(t) \phi_m(t) \phi_k(t) \, \mathrm{d}t 
 \\
&= 2(\lambda_k - \xi \alpha_1) c_k - 3\xi \alpha_2 \sum_{n=1}^{N} \sum_{m=1}^{N} c_n c_m \mathcal{T}_{nmk}, \quad k=1,2,\dots,N
\end{aligned}
\end{equation}

\vspace{0em} 

\begin{equation}
\label{eq:tensor_derivation_process}
\begin{aligned}
\mathcal{T}_{nmk} &= \left(\frac{2}{T}\right)^{\frac{3}{2}} \int_{0}^{T} \sin\left(\frac{n\pi t}{T}\right) \sin\left(\frac{m\pi t}{T}\right) \sin\left(\frac{k\pi t}{T}\right) \,\mathrm{d}t \\
&= \frac{1}{\pi\sqrt{2T}} \left( \frac{1-(-1)^{n-m+k}}{n-m+k} + \frac{1-(-1)^{m-n+k}}{m-n+k} + \frac{1-(-1)^{n+m-k}}{n+m-k} - \frac{1-(-1)^{n+m+k}}{n+m+k} \right)
\end{aligned}
\end{equation}
\vspace*{4pt}
\hrule 
\end{figure*}

Mathematically, the Galerkin method requires the projection of the residual onto each basis function $\phi_k(t)$ to be zero, i.e., $\langle R, \phi_k \rangle = 0$. By computing this inner product for $k = 1, 2, \dots, N$ and exploiting the orthogonality of the eigenfunctions, the projection is derived in \eqref{eq:algebraic_system} on the top of the next page, where $\mathcal{T}_{nmk}$ represents the $(n,m,k)$-th entry of the tensor $\mathcal{T}$ and is defined by the following integral of the triple product of basis functions
\begin{equation}
\label{eq:tensor_def}
    \mathcal{T}_{nmk} = \int_{0}^{T} \phi_n(t) \phi_m(t) \phi_k(t) \, \mathrm{d}t.
\end{equation}
Here, the tensor $\mathcal{T}$ is symmetric, i.e., the value of $\mathcal{T}_{nmk}$ is invariant under any permutation of indices. For the sinusoidal basis defined in \eqref{eq:eigenvalues-eigenfunctions}, the integral in \eqref{eq:tensor_def} admits a closed-form expression via the trigonometric product-to-sum identities, as shown in \eqref{eq:tensor_derivation_process} on the top of the next page.

By introducing the diagonal eigenvalue matrix $\mathbf{\Lambda} = \mathrm{diag}(\lambda_1, \dots, \lambda_N)$, the right-hand side of the $N$ residual equations derived in \eqref{eq:algebraic_system} can be expressed in a compact vector function form as
\begin{equation}
\label{eq:nonlinear_function}
    \mathbf{f}(\mathbf{c}) = 2(\mathbf{\Lambda} - \xi \alpha_1 \mathbf{I}_N) \mathbf{c} - 3\xi \alpha_2 \mathbf{q}(\mathbf{c}),
\end{equation}
where $\mathbf{A}= 2(\mathbf{\Lambda} - \xi \alpha_1 \mathbf{I}_N) \in \mathbb{R}^{N \times N}$ represents the linear component of the function. The vector $\mathbf{q}(\mathbf{c}) \in \mathbb{R}^N$ captures the nonlinear mode coupling arising from the drag force, where its $k$-th component is given by
\begin{equation}
    [\mathbf{q}(\mathbf{c})]_k = \sum_{n=1}^{N} \sum_{m=1}^{N} c_n c_m \mathcal{T}_{nmk} = \mathbf{c}^\mathrm{T} \mathbf{T}_k \mathbf{c}.
\end{equation}
Here, $\mathbf{T}_k \in \mathbb{R}^{N \times N}$ denotes the $k$-th frontal slice of the tensor defined in \eqref{eq:tensor_def}. While $\mathbf{c}=\mathbf{0}$ is a trivial solution to $\mathbf{f}(\mathbf{c})=\mathbf{0}$, the corresponding constant solution $v_N(t)=0$ is apparently not the optimal velocity profile. However, finding a non-trivial solution is challenging due to the existence of the nonlinear term $\mathbf{q}(\mathbf{c})$. To gain insight into the structure of the mode coupling, it is instructive to first examine the asymptotic regime where the static energy consumption and quadratic drag are both negligible, i.e., $\alpha_2 \to 0$ and $E_{\mathrm{s}} \to 0$. In this case, the zero-residual condition $\mathbf{f}(\mathbf{c})=\mathbf{0}$ simplifies to a linear and decoupled system as
\begin{equation}
\label{eq: 32}
    (\mathbf{\Lambda} - \xi \alpha_1 \mathbf{I}_N) \mathbf{c} = \mathbf{0}.
\end{equation}
Here, to facilitate a non-trivial solution $\mathbf{c}$, the matrix $(\mathbf{\Lambda} - \xi \alpha_1 \mathbf{I}_N)$ must be singular, indicating that $\xi=\lambda_n/\alpha_1$ holds for some $n$. For any such solution, the Dinkelbach auxiliary objective in \eqref{eq: 32} reduces to $-\xi E_{\mathrm{s}}$ that vanishes in the considered asymptotic regime, satisfying the zero-objective optimality condition of the Dinkelbach's method.

Under this circumstance, the equation in \eqref{eq: 32} has infinitely many non-trivial solutions whilst our task is to find the one that maximizes the EE. Recall that in \eqref{eq:new_opt_problem}, the optimal $\xi$ equals the maximal EE. Therefore, maximizing the EE $\xi$ corresponds to activating the fundamental mode $n=1$ because the eigenvalues $\lambda_n$ in \eqref{eq:eigenvalues-eigenfunctions} are monotonically decreasing. This activation implies that the fundamental coefficient $c_1$ becomes a free parameter, while all higher-order coefficients must vanish, i.e., $c_n = 0$ for $n \ge 2$. Although any feasible non-zero $c_1$ achieves the maximal EE in this asymptotic regime, maximizing its amplitude minimizes the CRB, thereby yielding the best sensing performance. By pushing $c_1$ to its upper limit determined by the physical constraints, namely the maximum speed in \eqref{eq:con_ineq} and the track length in \eqref{eq:con_dist}, the optimal velocity profile in this regime admits a closed-form expression, given as
\begin{equation}
\label{eq:analytic_v}
    v_{\alpha_2=0,E_{\mathrm{s}}=0}^*(t) = A^* \sin\left( \frac{\pi t}{T} \right),
\end{equation}
where $A^*=\min\left( V_{\max}, \frac{\pi L}{2T} \right)$. In this regard, the QoS constraint in \eqref{eq:QoS} is obtained as $\frac{{A^*}^2T^2}{2\pi^2} \ge \eta \frac{L^2}{4}$. Therefore, the above closed-form solution remains feasible provided that $\eta \le \frac{2{A^*}^2T^2}{\pi^2L^2}=\min(\frac{2V_{\max}^2T^2}{\pi^2L^2}, \frac{1}{2})$.

Notably, the analytic solution in \eqref{eq:analytic_v} provides a valuable insight, i.e, in the absence of nonlinear damping and static power consumption, the optimal energy-efficient velocity $v(t)$ reduces to a simple sinusoidal function. When $\alpha_2$ and $E_{\mathrm{s}}$ are non-negligible, or the QoS contraint is active, the optimization problem no longer admits a closed-form non-trivial solution. In this regard, we seek the coefficients $\mathbf{c}^*$ that minimize the residual norm $\|\mathbf{f}(\mathbf{c})\|_2$, and the corresponding general constrained optimization problem is formulated in the next section.

\section{Constrained Optimization Framework}
\label{sec:constrained_optimization}
Following the Galerkin discretization in Sec. \ref{sec:spectral_discretization}, the problem reduces to finding the spectral coefficients in vector $\mathbf{c}$ that minimize the residual norm $\|\mathbf{f}(\mathbf{c})\|_2$. However, a challenge arises because the optimization is subject to \eqref{eq:con_ineq} and \eqref{eq:con_dist}, which are constraints on the velocity profile $v(t)$. Since our optimization variables are the discrete coefficients $\mathbf{c}$, we must convert these time-domain inequalities into equivalent discrete algebraic constraints in terms of $\mathbf{c}$. This section introduces this conversion and subsequently applies the SCA algorithm to solve the formulated constrained problem.

\subsection{Reformulation of the Constraint  $v(t)\ge0$}
\label{sec:lower_bound}
To recast the non-negativity constraint $v(t) \ge 0$ into constraints in terms of the coefficient vector $\mathbf{c}$, we utilize the trigonometric identity $\sin(n\theta) = \sin(\theta) U_{n-1}(\cos\theta)$, where $U_{n-1}(\cdot)$ denotes the Chebyshev polynomial of the second kind of degree $(n-1)$~\cite{szeg1939orthogonal}. Consequently, the trial velocity function in \eqref{eq:v_approx} is expressed in the polynomial domain as
\begin{equation}
\label{eq: poly_form}
    v_N(x) = \sqrt{\frac{2}{T}} \sqrt{1-x^2}\sum_{n=1}^{N} c_n U_{n-1}(x),
\end{equation}
where the variable transformation is given by $x=\cos(\frac{\pi t}{T}) \in [-1,1]$ for $t \in [0, T]$. Since the term $\sqrt{1-x^2}$ is non-negative over the interval $[-1, 1]$, the condition $v_N(t) \ge 0$ is strictly equivalent to requiring the polynomial part $P(x) \triangleq \sum_{n=1}^{N} c_n U_{n-1}(x)$ to be non-negative for all $x \in [-1, 1]$. 

To facilitate the derivation, we rewrite $P(x)$ in its canonical form as
\begin{equation}
\label{eq: px}
    P(x) = \sum_{k=0}^{N-1} p_k x^k,
\end{equation}
where $\mathbf{p} = [p_0, \dots, p_{N-1}]^\mathrm{T}$ represents the algebraic coefficients. The relationship between $\mathbf{p}$ and $\mathbf{c}$ is derived via the explicit expansion of $U_{n-1}(x)$, which is given by~\cite{szeg1939orthogonal}
\begin{equation}
    U_{n-1}(x) = \sum_{j=0}^{\lfloor \frac{n-1}{2} \rfloor} (-1)^j \binom{n-1-j}{j} (2x)^{n-1-2j}.
\end{equation}

By substituting this expansion into $P(x)$ and grouping terms by powers of $x$, the linear mapping can be compactly expressed as $\mathbf{p} = \mathbf{M} \mathbf{c}$. The non-zero entries of the transformation matrix $\mathbf{M}\in \mathbb{R}^{N \times N}$, denoted by $M_{k+1,n}$, is given by
\begin{equation}
\label{eq: M}
M_{k+1,n} = (-1)^{\frac{n-1-k}{2}} \binom{\frac{n-1+k}{2}}{\frac{n-1-k}{2}} 2^{k},
\end{equation}
provided that both $n-1 \ge k$ and $(n-1-k) \bmod 2 = 0$ and otherwise, $M_{k+1,n} = 0$. Now, we can invoke the theorem of Markov-Lukács to characterize the non-negativity of $P(x)$ exactly via SOS decomposition.

\begin{lemma}[Theorem of Markov-Lukács~\cite{szeg1939orthogonal}]
\label{lemma:lukacs}
    Let $P(x)$ be a polynomial of degree $D$. Then, $P(x) \ge 0$ for all $x \in [-1, 1]$ if and only if $P(x)$ can be expressed in the following forms:
    \begin{itemize}
        \item If $D$ is even, then $P(x) = s_1(x) + (1-x^2)s_2(x)$, where $\deg(s_1) \le D$ and $\deg(s_2) \le D-2$.
        \item If $D$ is odd, then $P(x) = (1+x)s_1(x) + (1-x)s_2(x)$, where $\deg(s_1) \le D-1$ and $\deg(s_2) \le D-1$,
    \end{itemize}
where $s_1(x)$ and $s_2(x)$ are SOS polynomials.
\end{lemma}

To obtain computationally tractable constraints from the SOS polynomials, we invoke the Gram matrix representation, which allows us to parameterize any SOS polynomial as a quadratic form. This compact representation is summarized in the following lemma.

\begin{lemma}[Gram Matrix Representation of SOS Polynomials~\cite{10383828, roh2006discrete}]
\label{lemma:gram_matrix}
    A polynomial $s(x)$ of degree $2d$ is an SOS polynomial if and only if there exists a PSD matrix $\mathbf{Q} \in \mathbb{S}^{d+1}$ such that
    \begin{equation}
        s(x) = \mathbf{z}_d^\mathrm{T}(x) \mathbf{Q} \mathbf{z}_d(x),
    \end{equation}
    where $\mathbf{z}_d(x) = [1, x, \dots, x^d]^T$ denotes the standard monomial basis vector with order $d$.
\end{lemma}

We take the case that the number of modes $N$ is odd as an example, and similar derivations apply for even values of $N$. Consequently, the degree of $P(x)$, $N-1$, is even, corresponding to the first case in Lemma \ref{lemma:lukacs}. Applying the representation in Lemma \ref{lemma:gram_matrix} to $P(x)$ implies the existence of two auxiliary PSD matrices $\mathbf{Q}_1 \in \mathbb{S}^{\frac{N+1}{2}}$ and $\mathbf{Q}_2 \in \mathbb{S}^{\frac{N-1}{2}}$, such that
\begin{equation}
\label{eq:sos_poly_relation}
\begin{aligned}
    \sum_{k=0}^{N-1} p_k x^k &= \mathbf{z}_{\frac{N-1}{2}}^\mathrm{T}(x) \mathbf{Q}_1 \mathbf{z}_{\frac{N-1}{2}}(x) \\
    &\quad  + (1-x^2) \mathbf{z}_{\frac{N-3}{2}}^\mathrm{T}(x) \mathbf{Q}_2 \mathbf{z}_{\frac{N-3}{2}}(x).
\end{aligned}
\end{equation}

To enforce this algebraically, we define the linear operator $\mathcal{H}_k(\mathbf{Q})$ in terms of $\mathbf{Q}\in \mathbb{S}^{N}$, which returns the coefficient of $x^k$ in the quadratic form $\mathbf{z}_{N-1}^\mathrm{T} \mathbf{Q} \mathbf{z}_{N-1}$ as
\begin{equation}
    \mathcal{H}_k(\mathbf{Q}) = \sum_{\substack{1 \le i,j \le N\\ i \le j,\, i+j-2=k}} 2Q_{i,j}.
\end{equation}

Utilizing this operator and matching the coefficients on both sides of \eqref{eq:sos_poly_relation} yields a set of linear equality constraints as
\begin{equation}
\label{eq:coeff_matching_explicit}
    p_k = \mathcal{H}_k(\mathbf{Q}_1) + \mathcal{H}_k(\mathbf{Q}_2) - \mathcal{H}_{k-2}(\mathbf{Q}_2), \quad k = 0, \dots, N-1,
\end{equation}
with the convention that $\mathcal{H}_{k}(\cdot) = 0$ if $k < 0$. 

In summary, the linear equality constraints derived in \eqref{eq:coeff_matching_explicit}, combined with the PSD condition on $\mathbf{Q}_1$ and $\mathbf{Q}_2$, provide a necessary and sufficient condition for $P(x) \ge 0$ on $[-1, 1]$. Recalling the equivalence established in \eqref{eq: poly_form}, satisfying these algebraic constraints guarantees the physical non-negativity constraint $v(t) \ge 0$ over the entire sensing interval $[0, T]$.

\subsection{Reformulation of the Constraint $v(t)\le V_{\max}$}
\label{sec:upper_bound}
This section reformulates the maximum velocity limit into another set of constraints in terms of $\mathbf{c}$. Leveraging the polynomial structure established in \eqref{eq: poly_form}, the upper bound constraint $v(t)\le V_{\max}$ can be rewritten as
\begin{equation}
\label{eq: ori_upper}
    \sqrt{\frac{2}{T}} \sqrt{1-x^2} P(x) \le V_{\max}, \quad \forall x \in [-1, 1],
\end{equation}
where $P(x)$ is the polynomial part of the velocity derived in \eqref{eq: px}. However, directly applying the SOS decomposition in \eqref{eq: ori_upper} is intractable due to the non-polynomial term $\sqrt{1-x^2}$. While squaring both sides eliminates the non-polynomial factor, it would introduce highly non-convex quadratic constraints in terms of $\mathbf{c}$. To streamline the formulation, we introduce a tight polynomial sufficient condition. By exploiting the algebraic inequality $\sqrt{1-x^2} \le 1 - \frac{1}{2}x^2$ for all $x \in [-1, 1]$, and noting that $P(x) \ge 0$, it follows that $\sqrt{1-x^2} P(x) \le (1 - \frac{1}{2}x^2) P(x)$. Therefore, \eqref{eq: ori_upper} can be satisfied by requiring the polynomial $G(x)$ to be non-negative, defined as
\begin{equation}
\label{eq: 43}
    G(x) \triangleq \sqrt{\frac{T}{2}}V_{\max} - (1 - \frac{1}{2}x^2)P(x) \ge 0, \quad \forall x \in [-1, 1].
\end{equation}


Following the methodology in Sec. \ref{sec:lower_bound}, we express $G(x)$ in the canonical form as $G(x) = \sum_{k=0}^{N+1} g_k x^k$. Note that the degree of $G(x)$ is $N+1$ due to the multiplication of $1-0.5x^2$ and $P(x)$. By expanding the polynomial terms, we derive an affine mapping between $\mathbf{g}=[g_0,g_1,\dots,g_{N+1}]^\mathrm{T}$ and $\mathbf{p}$ as
\begin{equation}
    \mathbf{g} = \mathbf{F}\mathbf{p}+\mathbf{b},
\end{equation}
where the transformation matrix $\mathbf{F}$ and bias vector $\mathbf{b}$ are defined as
\begin{equation}
\label{eq: Fb}
    \mathbf{F} = \left[
\begin{array}{c}
  \mathbf{0}_{2 \times N} \\
  \frac{1}{2} \mathbf{I}_{N}
\end{array}
\right] - 
\left[
\begin{array}{c}
    \mathbf{I}_{N} \\
  \mathbf{0}_{2 \times N}
\end{array}
\right],
\quad \mathbf{b} = \left[
\begin{array}{c}
  \sqrt{\frac{T}{2}}V_{\max} \\
  \mathbf{0}_{(N+1)\times 1}
\end{array}
\right],
\end{equation}
respectively. Therefore, substituting $\mathbf{p} = \mathbf{M} \mathbf{c}$, $\mathbf{g}$ can be linearly represented by $\mathbf{c}$ via $\mathbf{g} = \mathbf{F}\mathbf{M}\mathbf{c}+\mathbf{b}$. With $G(x)$ fully characterized by $\mathbf{g}$, we apply Lemma \ref{lemma:lukacs} again to enforce its non-negativity. Recall that we have assumed that $N$ is odd in Sec. \ref{sec:lower_bound}, the degree of $G(x)$ is even. Thus, there exist two auxiliary PSD matrices $\mathbf{Y}_1 \in \mathbb{S}^{\frac{N+3}{2}}$ and $\mathbf{Y}_2 \in \mathbb{S}^{\frac{N+1}{2}}$ such that
\begin{equation}
\begin{aligned}
    G(x) &= \mathbf{z}_{\frac{N+1}{2}}^\mathrm{T}(x) \mathbf{Y}_1 \mathbf{z}_{\frac{N+1}{2}}(x) \\
    &\quad+ (1-x^2) \mathbf{z}_{\frac{N-1}{2}}^\mathrm{T}(x) \mathbf{Y}_2 \mathbf{z}_{\frac{N-1}{2}}(x).
\end{aligned}
\end{equation}

Finally, matching the coefficients on both sides leads to the affine constraints for the optimization problem as
\begin{equation}
\label{eq:g_const}
    g_k = \mathcal{H}_k(\mathbf{Y}_1) + \mathcal{H}_k(\mathbf{Y}_2) - \mathcal{H}_{k-2}(\mathbf{Y}_2), \quad k=0, 1, \dots, N+1.
\end{equation}



\subsection{Optimization Problem Formulation}
In Secs. \ref{sec:lower_bound} and \ref{sec:upper_bound}, the velocity constraints \eqref{eq:con_ineq} have been transformed into the SOS-based constraints in terms of $\mathbf{c}$. The remaining constraints that required to be reformulated are \eqref{eq:con_dist} and \eqref{eq:QoS}. First, substituting the trial solution in \eqref{eq:v_approx} into \eqref{eq:con_dist} yields
\begin{equation}
\label{eq:inner_prod_con}
    \sum_{n=1}^{N} c_n \int_{0}^{T} \phi_n(\tau) \, \mathrm{d}\tau = \bm{\beta}^\mathrm{T}\mathbf{c}
    \le L,
\end{equation}
where $\bm{\beta}$ is a column vector of length $N$ whose $n$-th entry, denoted as $\beta_n$, is given by
\begin{equation}
    \beta_n = \int_{0}^{T} \phi_n(\tau) \, \mathrm{d}\tau = \frac{\sqrt{2T}\left[1-(-1)^n\right]}{n\pi}.
\end{equation}

Second, by substituting \eqref{eq:v_approx} into \eqref{eq: kernel_rep_v}, we obtain the spatial variance as $\mathbf{c}^\mathrm{T}\mathbf{\Lambda}\mathbf{c}$. Then, \eqref{eq:QoS} is reformulated as
\begin{equation}
\label{eq:qos_con}
\mathbf{c}^\mathrm{T}\mathbf{\Lambda}\mathbf{c} \ge \eta \frac{L^2}{4}.
\end{equation}

Consequently, the original variational problem can be reformulated into a constrained nonlinear least-squares problem as
\begin{subequations}
\label{eq:final_optimization}
\begin{align}
    \min_{\boldsymbol{\Theta}} \quad & \Psi(\mathbf{c}) = \frac{1}{2} \|\mathbf{f}(\mathbf{c})\|_2^2 \label{eq:obj_func}
    \\
    \text{s.t.} \quad 
    & \mathbf{p} = \mathbf{M}\mathbf{c}, \label{eq:pc}\\
    & \mathbf{g} = \mathbf{F}\mathbf{M}\mathbf{c}+\mathbf{b}, \label{eq:gc}\\
    & \eqref{eq:coeff_matching_explicit}, \eqref{eq:g_const},\eqref{eq:inner_prod_con},\eqref{eq:qos_con} \label{eq:con_match} \\
    & \mathbf{Q}_1, \mathbf{Q}_2, \mathbf{Y}_1, \mathbf{Y}_2 \succeq 0, \label{eq:con_psd}
\end{align}
\end{subequations}
where the optimization variable $\boldsymbol{\Theta}$ is defined as $\{\mathbf{c}, \mathbf{Q}_1, \mathbf{Q}_2, \mathbf{Y}_1, \mathbf{Y}_2\}$. To address the non-convexity of the objective function and the QoS constraint in \eqref{eq:qos_con} while leveraging the convexity of the other constraints, we employ the SCA algorithm to solve the problem iteratively. First, exploiting the nonlinear least-squares structure of the objective function, a strictly convex surrogate objective is constructed in the $k$-th iteration via the Gauss-Newton approximation~\cite{gratton2007approximate}. By linearizing the residual vector $\mathbf{f}(\mathbf{c})$ at the current local point $\mathbf{c}^{(k)}$, the surrogate objective is given by
\begin{equation}
\label{eq: surrogate_func}
\tilde{\Psi}(\mathbf{c};\mathbf{c}^{(k)}) = \frac{1}{2} \| \mathbf{f}(\mathbf{c}^{(k)}) + \mathbf{J}(\mathbf{c}^{(k)}) (\mathbf{c}-\mathbf{c}^{(k)}) \|_2^2,
\end{equation}
where $\mathbf{J}(\mathbf{c}^{(k)})$ is the Jacobian matrix of $\mathbf{f}(\mathbf{c})$ evaluated at $\mathbf{c}^{(k)}$, given by
\begin{equation}
\label{eq: Jacobian_ck}
    \mathbf{J}(\mathbf{c}^{(k)}) = \mathbf{A} - 6\xi \alpha_2 \mathbf{K}(\mathbf{c}^{(k)}),
\end{equation}
where the $(k,j)$-th entry of the coupling matrix $\mathbf{K}(\mathbf{c})$ is defined by $\sum_{n=1}^N c_n \mathcal{T}_{njk}$. The detailed derivation of \eqref{eq: Jacobian_ck} is provided in Appendix \ref{sec:app3}. It can be observed from \eqref{eq: surrogate_func} that the Gauss-Newton method naturally produces a convex quadratic objective, eliminating the need for manual parameter tuning. 

\begin{algorithm}[!t]
\small
\caption{Proposed Velocity Profile Optimization}
\label{alg:velocity_opt}
\begin{algorithmic}[1]
\Require Parameters $P_{\mathrm{s}}, T, L, V_{\max}, \alpha_1, \alpha_2, N, \eta$; Convergence tolerances $\epsilon_{\text{out}}$ and $\epsilon_{\text{in}}$ for the outer and inner loops, respectively.
\Ensure Optimal spectral coefficients $\mathbf{c}^*$ and corresponding velocity profile $v^*(t)$.
\State Define $\mathbf{M}$, $\mathbf{F}$, and $\mathbf{b}$ according to \eqref{eq: M} and \eqref{eq: Fb}, respectively. Initialize the outer index $i=0$ and the starting point $\hat{\mathbf{c}}^{(0)} = [c_1^{(0)}, 0, \dots, 0]^\mathrm{T}$, where $c_1^{(0)} = \min(V_{\max}, \frac{\pi L}{2T})\sqrt{T/2}$.
\Repeat
    \State Initialize coefficients $\mathbf{c}^{(0)} = \hat{\mathbf{c}}^{(i)}$ and inner index $k=0$.
    \Repeat 
        \State Construct the surrogate function $\tilde{\Psi}(\mathbf{c};\mathbf{c}^{(k)})$ via \eqref{eq: surrogate_func}.
        \State Update $\mathbf{c}^{(k+1)}$ via solving the convex problem in \eqref{eq:final_optimization}.
        \State Update $k \gets k+1$.
    \Until{$\|\mathbf{c}^{(k)} - \mathbf{c}^{(k-1)}\|_2 \le \epsilon_{\text{in}}$}.
    \State Set the optimal coefficients $\hat{\mathbf{c}}^{(i)} = \mathbf{c}^{(k)}$.
    \State Reconstruct the velocity profile $v^{(i)}(t)$ via \eqref{eq:eigenvalues-eigenfunctions} and \eqref{eq:v_approx}. 
    \State Calculate $\mathcal{V}(v^{(i)})$ and $\mathcal{E}(v^{(i)})$ via \eqref{eq: 1}, \eqref{eq: var}, and \eqref{eq: energy_consumption}.
    \State Update $\xi^{(i+1)} = \mathcal{V}(v^{(i)})/\mathcal{E}(v^{(i)})$ and $i \gets i+1$.
\Until{$|\xi^{(i)} - \xi^{(i-1)}| \le \epsilon_{\text{out}}$}. 
\end{algorithmic}
\end{algorithm}

In addition to the objective function, the QoS constraint in \eqref{eq:qos_con} is non-convex since the quadratic term $\mathbf{c}^\mathrm{T}\mathbf{\Lambda}\mathbf{c}$ is convex. To tackle this, we replace it with its global affine lower bound via the first-order Taylor expansion at the current local point $\mathbf{c}^{(k)}$, yielding
\begin{equation}
\label{eq:qos_linear}
    \mathbf{c}^{(k)\mathrm{T}} \mathbf{\Lambda} \mathbf{c}^{(k)} + 2\mathbf{c}^{(k)\mathrm{T}} \mathbf{\Lambda} (\mathbf{c} - \mathbf{c}^{(k)}) \ge \eta \frac{L^2}{4}.
\end{equation}

Consequently, at each iteration, the non-convex problem is approximated by minimizing the convex surrogate function \eqref{eq: surrogate_func} subject to the convexified constraint \eqref{eq:qos_linear} alongside the remaining constraints in \eqref{eq:final_optimization}. Therefore, this subproblem can be solved efficiently by using standard convex optimization tools like CVX~\cite{boyd2004convex,grant2014cvx}. Combined with the Dinkelbach's algorithm in Sec. \ref{sec: dinklebach}, the overall optimization procedure is summarized in \textbf{Algorithm \ref{alg:velocity_opt}}.

\section{Simulations}
\label{sec:simulation}
In this section, we conduct numerical simulations to evaluate the performance of the proposed energy-efficient velocity profile optimization framework for MA-assisted 1-D DoA estimation.

\subsection{Simulation Setup}
Unless otherwise specified, the simulations are conducted with the following parameters. The MA system moves along a linear track of length $L=4$ m over a sensing interval of $T=1$ s. The mechanical properties of the MA are characterized by a mass of $m_{\mathrm{a}} = 0.1$ kg, a linear damping coefficient of $\alpha_1 = 0.2$ kg/s, and a nonlinear drag coefficient of $\alpha_2 = 0.1$ kg/m. The maximum speed of the MA is $V_{\max}=10$ m/s. The static power consumption $P_{\mathrm{s}}$ is set to $0.1$ W. The fraction $\eta$ is set to $0.1$.

To systematically evaluate the effectiveness of the proposed velocity profile optimization and elucidate the fundamental trade-offs, we select the following benchmarks based on a 2-D logical framework, i.e., the velocity modeling space and the optimization objective.

\begin{enumerate}
    \item \textit{Uniform Velocity [Performance-Oriented]} (labeled as \textbf{[Perf] Scalar}):
    This benchmark takes a sensing performance-oriented perspective\footnote{Sensing EE optimization in the scalar space yields a trivial zero-speed solution with an infinite CRB, and is therefore excluded from our evaluations.} to maximize spatial coverage by moving the antenna at the maximum allowable constant speed, i.e., $v(t) = \min(V_{\max}, L/T)$.

    \item \textit{Binary [Performance-Oriented]} (labeled as \textbf{[Perf] Binary}): 
    In this benchmark, $v(t)$ is discretized into an $M$-dimensional vector $\mathbf{v}$. To maximize the sensing performance, the MA adopts a binary switching strategy~\cite{ma2025movable}, where the velocity in the $m$-th time slot takes only the extreme values, i.e., $v_m^* \in \{0, V_{\max}\}$.

    \item \textit{Sinusoidal Velocity Profile [EE-Oriented]} (labeled as \textbf{[EE] Sinusoidal}): 
    This scheme utilizes the closed-form solution derived in \eqref{eq:analytic_v}. Comparing against this solution visualizes the capability of the proposed algorithm to adaptively reshape the trajectory in the presence of strong nonlinear aerodynamic drag.
    
    \item \textit{Trapezoidal Profile [EE-Oriented]} (labeled as \textbf{[EE] Trapezoidal}): 
    As a widely adopted motion profile~\cite{7260323}, the trapezoidal profile consists of three sequential phases: constant acceleration, constant velocity, and constant deceleration. With the acceleration and deceleration periods restricted to an identical duration, this single temporal parameter is optimized via a 1-D search to maximize the sensing EE.
\end{enumerate}

\subsection{Convergence and Physical Interpretation}
\label{subsec:convergence}
In this subsection, we validate the numerical stability and consistency of the proposed algorithm and elucidate the energy-efficient velocity profile.

\subsubsection{Convergence Analysis}
\begin{figure}[!h]
    \centering
    \subfloat[Impact of the truncation order $N$.]{
        \includegraphics[width=0.45\columnwidth]{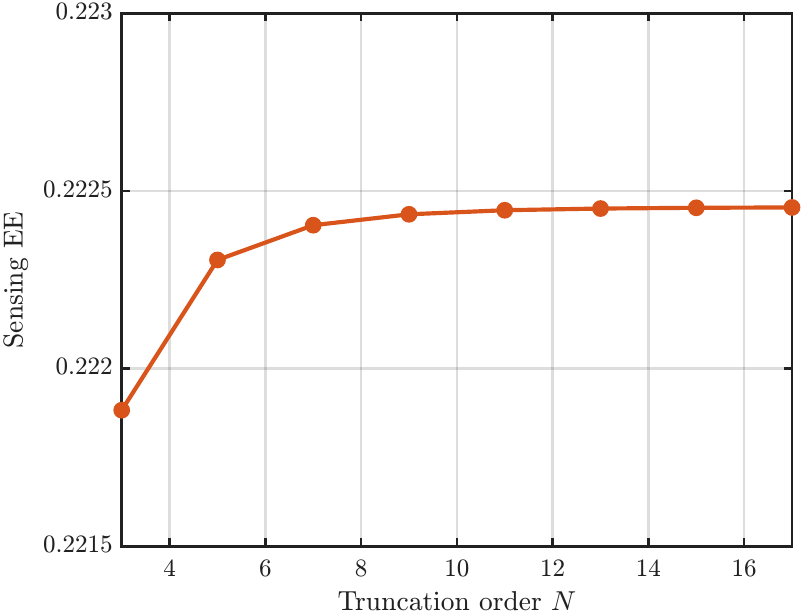}
        \label{fig:resolution_N}
    }
    \hfil
    \subfloat[Impact of the QoS requirement $\eta$.]{
        \includegraphics[width=0.45\columnwidth]{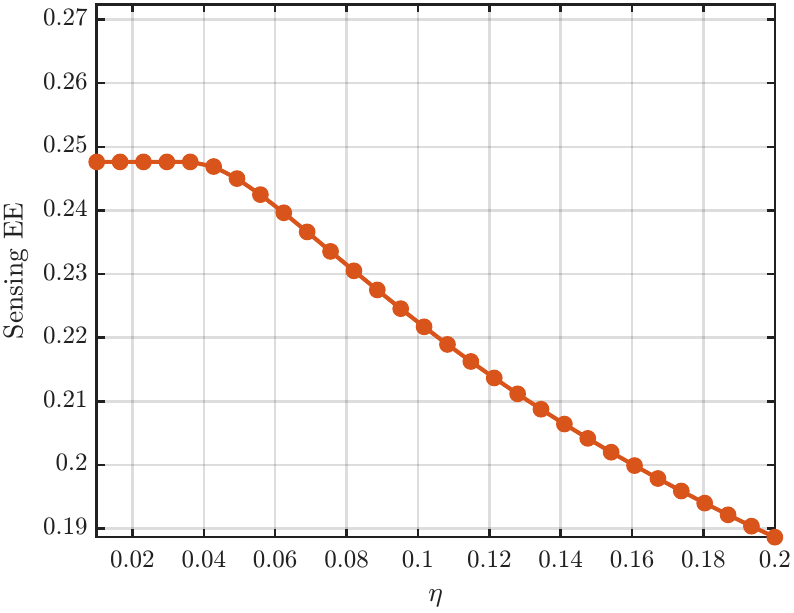}
        \label{fig:resolution_eta}
    }
    \caption{Impact of the truncation order and sensing QoS requirement on the sensing EE.}
    \label{fig:resolution}
\end{figure}

We first examine the impact of the truncation parameter $N$ in~\eqref{eq:v_approx} on the achieved sensing EE for the proposed scheme. As depicted in Fig.~\ref{fig:resolution}(a), the sensing EE performance saturates rapidly, reaching its peak at a relatively small order of $N=11$. Therefore, we set the truncation order to $N=11$ for all subsequent simulations.

We further investigate the impact of the sensing QoS requirement $\eta$ in Fig.~\ref{fig:resolution}(b). It can be observed that the sensing EE remains nearly unchanged when $\eta$ is relatively small, since the optimized velocity profile already satisfies the required sensing quality. As $\eta$ increases, the sensing EE gradually decreases because a stricter sensing requirement forces the MA to achieve a larger spatial variance, which requires more aggressive antenna movement and consequently incurs higher mechanical energy consumption.

\begin{figure}[!ht]
    \centering
    \subfloat[$\alpha_2 = 1 \times 10^{-3}$]{
        \includegraphics[width=0.45\columnwidth]{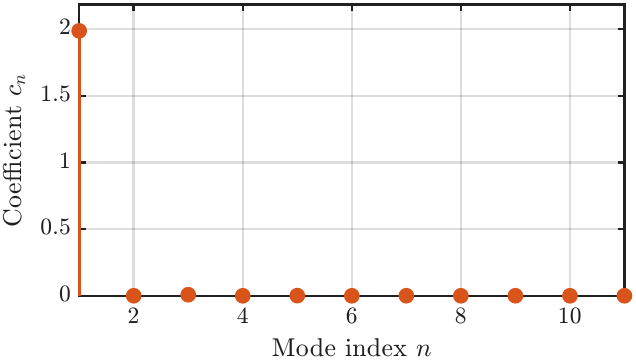}
        \label{fig:coeff_baseline}
    }
    \hfil 
    \subfloat[$\alpha_2 = 1 \times 10^{-2}$]{
        \includegraphics[width=0.45\columnwidth]{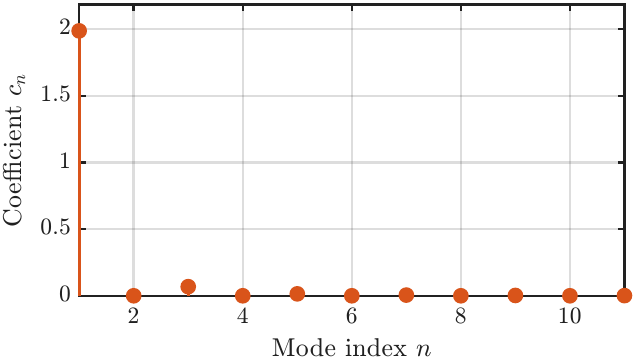}
        \label{fig:coeff_alpha1}
    } 
    \\
    \subfloat[$\alpha_2 = 1 \times 10^{-1}$]{
        \includegraphics[width=0.45\columnwidth]{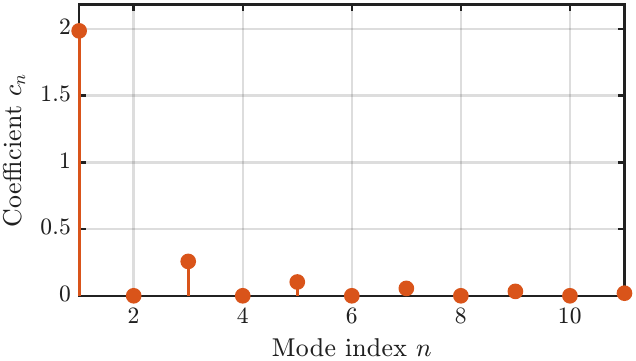}
        \label{fig:coeff_alpha2}
    }
    \hfil
    \subfloat[$\alpha_2 = 1 \times 10^{0}$]{
    \includegraphics[width=0.45\columnwidth]{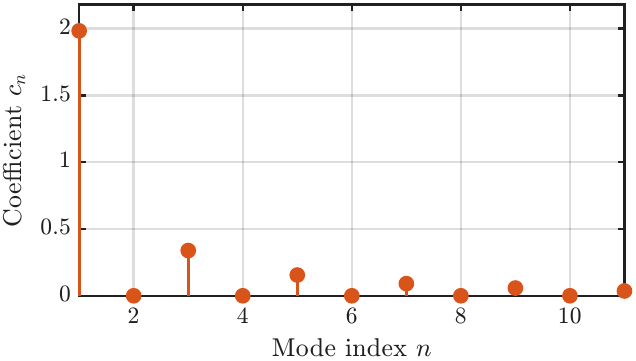}
        \label{fig:coeff_alpha3}
    }
    \caption{Comparison of the optimized coefficients $\mathbf{c}$ under different $\alpha_2$.}
    \label{fig:coeff_comparison}
\end{figure}

Next, we verify the mathematical consistency of the proposed optimization framework with the `[EE] sinusoidal' baseline in the asymptotic regime where the nonlinear aerodynamic drag is negligible, i.e., a small $\alpha_2$. Fig.~\ref{fig:coeff_comparison} illustrates the distribution of the optimized spectral coefficients in vector $\mathbf{c}$ under varying nonlinear drag weights $\alpha_2 \in \{10^{-3}, 10^{-2}, 10^{-1}, 10^{0}\}$. As observed in Fig.~\ref{fig:coeff_comparison}(a), when $\alpha_2$ is sufficiently small, only the first fundamental mode is activated while the higher-order coefficients remain strictly zero. This observation aligns perfectly with the theoretical analysis after \eqref{eq: 32} in Sec.~\ref{sec:spectral_discretization}.

\begin{figure}[!h]
    \centering    \includegraphics[width=0.35\textwidth]{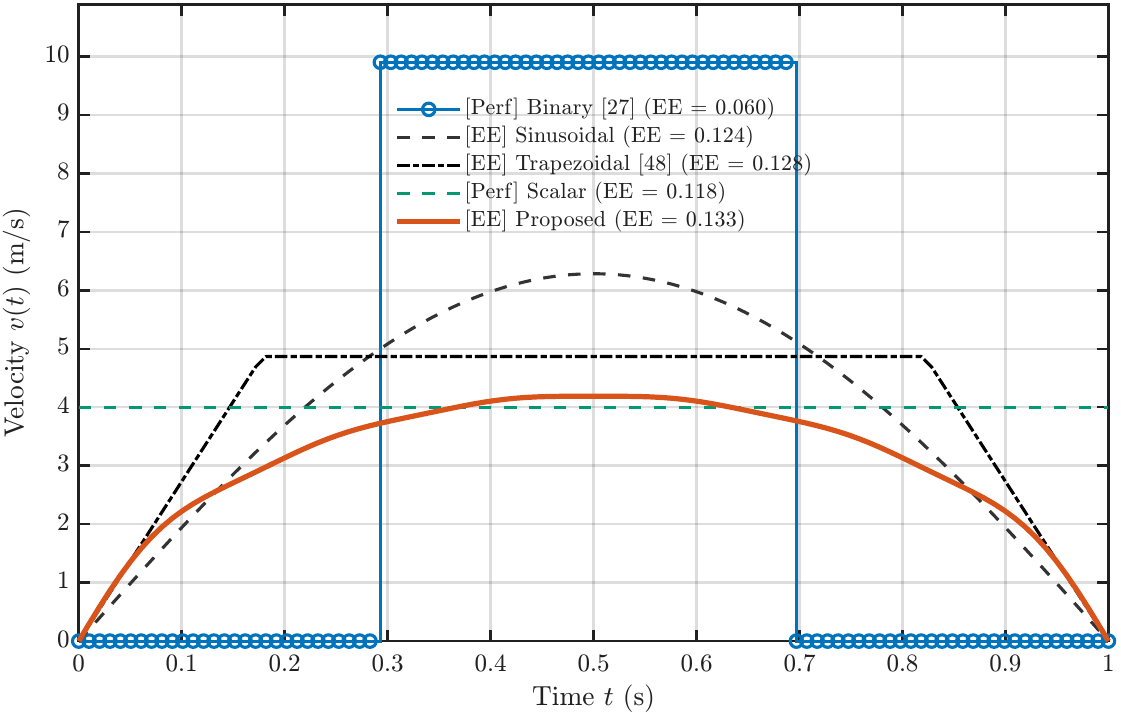} 
    \caption{Comparison of optimized velocity profiles under different methods.}
    \label{fig:velocity_profiles_comparison}
\end{figure}

As $\alpha_2$ gradually increases from $10^{-2}$ to $10^{0}$, as shown in Figs.~\ref{fig:coeff_comparison}(b)-(d), the first mode remains dominant, but higher-order spectral coefficients are progressively activated. Physically, this indicates that as the nonlinear aerodynamic drag becomes non-negligible and even dominant, the proposed method adaptively shapes the velocity profile by incorporating higher-frequency components to suppress the excessive energy consumption induced by the nonlinear drag, thereby preserving the sensing EE. Overall, these results demonstrate the strong adaptability of the proposed framework. Furthermore, they suggest that in scenarios where the nonlinear drag is negligible, the computationally efficient sinusoidal profile can be directly adopted, eliminating the need for iterative numerical optimization.

\subsubsection{Physical Interpretation of Velocity Profiles}
To provide deeper insights into the underlying mechanisms that boost the sensing EE, Fig.~\ref{fig:velocity_profiles_comparison} visualizes the optimized velocity profiles alongside their achieved sensing EE values. Notably, the `[EE] Proposed' method achieves the highest sensing EE of 0.133 $1/\text{rad}^2/\text{Joule}$ with a smooth profile. This can be explained by Fig.~\ref{fig:coeff_comparison}, which shows that low-frequency components completely dominate, allowing the optimal profile to naturally avoid non-smooth high-frequency oscillations. While the profile of the `[EE] Sinusoidal' baseline (EE = 0.124) is also smooth and theoretically optimal under zero-drag conditions, its performance inevitably degrades in the presence of strong aerodynamic drag coefficient $\alpha_2$. Conversely, the remaining baselines suffer from severe physical discontinuities. Specifically, the `[EE] Trapezoidal' (EE = 0.128), `[Perf] Scalar' (EE = 0.118), and `[Perf] Binary' (EE = 0.060) schemes all introduce abrupt velocity transitions or infinite corner accelerations. Consequently, the rigid or greedy nature of these strategies inevitably leads to severe mechanical energy consumption and suboptimal sensing EE.

\begin{figure}[!h]
    \centering    \includegraphics[width=0.33\textwidth]{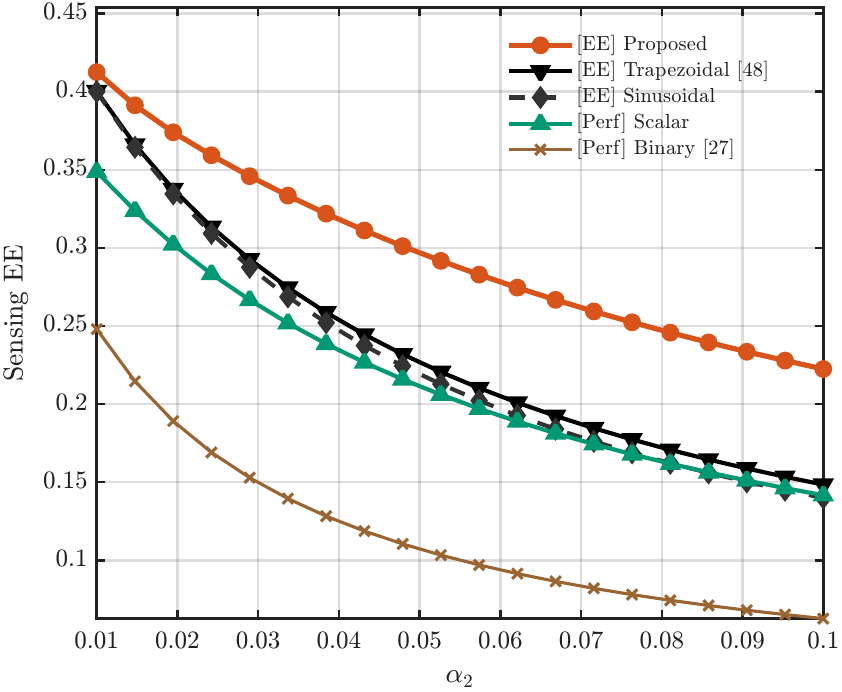} 
    \caption{Sensing EE performance versus the nonlinear drag coefficient $\alpha_2$.}
    \label{fig:ee_alpha2}
\end{figure}

To further demonstrate the adaptability of the proposed framework, Fig.~\ref{fig:ee_alpha2} compares the sensing EE versus the nonlinear drag coefficient $\alpha_2$ across different schemes. When $\alpha_2$ is small, the sensing EE of the `[EE] Proposed' method closely matches that of the `[EE] Sinusoidal' baseline, which again verifies the optimality of the latter under zero-drag conditions. As $\alpha_2$ increases, the `[EE] Proposed' method increasingly outperforms both the `[EE] Sinusoidal' and `[EE] Trapezoidal' baselines. This widening performance gap highlights the inherent flexibility of the spectral representation, which leverages multi-mode DoFs to dynamically sculpt the velocity profile against strong nonlinear drag. Meanwhile, the curves for the `[Perf] Scalar' and `[Perf] Binary' baselines exhibit the worst performance. This behavior is intuitive, as these performance-oriented designs entirely neglect the mechanical energy consumption, leading to low overall sensing EE.

\subsection{Impact of System Configurations}
\label{subsec:system_impact}
\begin{figure}[!h]
	\centering
	\includegraphics[width=0.33\textwidth]{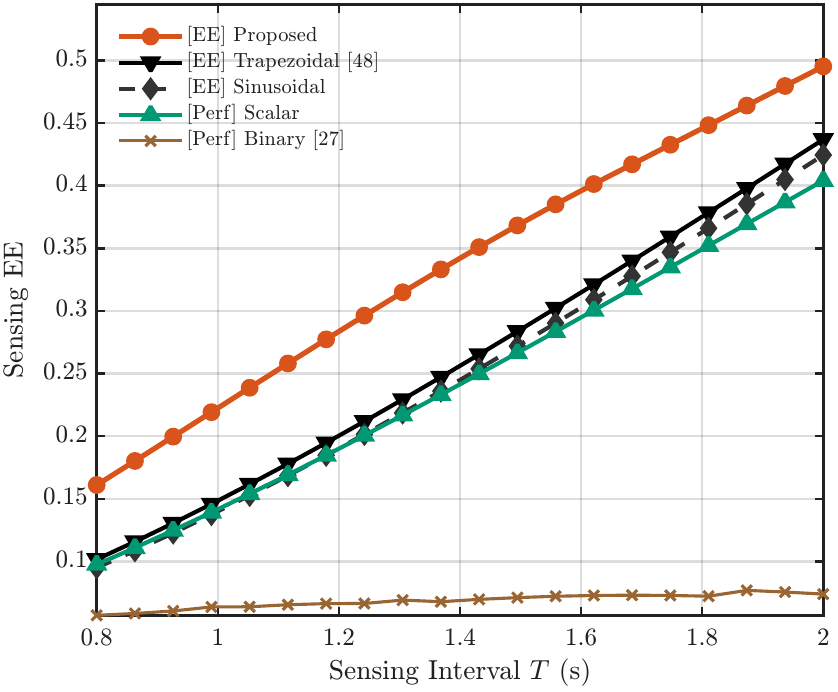}
	\caption{Sensing EE performance versus the sensing interval $T$.} 
	\label{fig:EE_time}
\end{figure}

\begin{figure}[!h]
	\centering
	\includegraphics[width=0.33\textwidth]{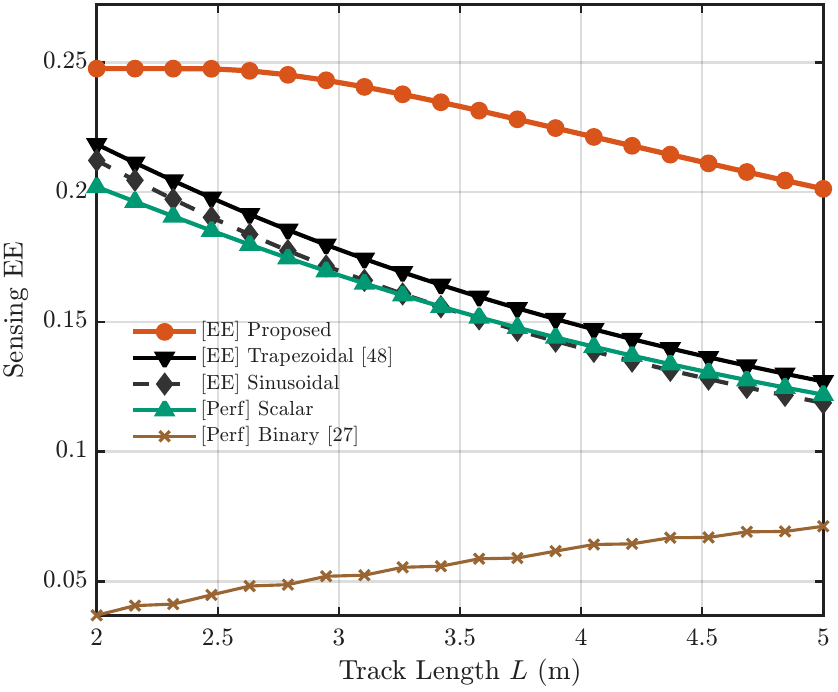}
	\caption{Sensing EE performance versus the track length $L$.} 
	\label{fig:EE_length}
\end{figure}

This subsection evaluates the sensing EE in different system configurations by the sensing interval $T$ and track length $L$. The simulation results are presented in Fig.~\ref{fig:EE_time} and Fig.~\ref{fig:EE_length}, respectively.

Fig.~\ref{fig:EE_time} depicts the sensing EE versus the sensing interval $T$. It is observed that the sensing EE of all schemes monotonically increases with $T$ except for the `[Perf] Binary' method. This trend is physically intuitive: for a fixed track length $L$, extending the sensing interval $T$ allows the MA to travel at a lower average velocity. As shown in \eqref{eq: energy_consumption}, the aerodynamic drag energy scales cubically with the instantaneous speed, this velocity reduction yields substantial energy savings that far outweigh the energy consumed by the extended operation time. Notably, the `[EE] Proposed' algorithm consistently outperforms all baseline schemes. In comparison, the `[EE] Trapezoidal' and `[EE] Sinusoidal' baselines achieve only moderate EE, as their performances are fundamentally bottlenecked by rigid profile structures and the idealized zero-drag assumption, respectively. In contrast, the two performance-oriented schemes yield the lowest sensing EE. By greedily maximizing the spatial variance at high velocities, they expose the critical inefficiency of neglecting mechanical energy consumption during the system design. Specifically, the `[Perf] Binary' curve remains stagnant because it always sprints at the maximum velocity regardless of the sensing interval $T$, thereby constantly suffering from the peak drag penalty.

Next, Fig.~\ref{fig:EE_length} illustrates the impact of the track length $L$ on the sensing EE. With the exception of the `[Perf] Binary' benchmark, the sensing EE of all evaluated schemes decreases monotonically as $L$ extends. Covering a longer distance within a fixed sensing interval $T$ inevitably demands higher average velocities. Consequently, the severe cubic energy penalty induced by the aerodynamic drag rapidly overshadows the quadratic spatial diversity gain. Nevertheless, the `[EE] Proposed' scheme still outperforms all  baselines. Interestingly, the `[Perf] Binary' scheme exhibits a contrasting increasing trend. This behavior emerges because its corresponding energy consumption grows only linearly with $L$, which is strictly outpaced by the quadratic expansion of the spatial variance.

\section{Conclusion}
\label{sec:conclusion}
This paper shifted the MA design paradigm for 1-D DoA estimation from conventional motion models to time-varying velocity profile optimization to maximize the sensing EE. By bridging the calculus of variations and spectral theory, we derived a closed-form optimal solution for the regime with linear drag. For the general scenarios with significant nonlinear drag, we leveraged SOS decomposition to transform the time-domain constraints into a tractable algebraic formulation, and proposed a two-layer algorithm synthesizing Dinkelbach’s method and SCA. The numerical results have verified that the proposed framework significantly outperforms conventional benchmarks. Critically, our analysis revealed that the optimal velocity profile is inherently smooth and dominated by low-frequency spectral modes, achieving a good trade-off between the sensing performance and mechanical energy consumption. Future work may investigate the extension of this velocity profile optimization framework to multi-MA cooperative systems or ISAC scenarios.

\begin{appendices}
\section{Proof of Proposition 1}
\label{sec:app1}
Recall that the variance in \eqref{eq: var} is given by
\begin{equation}
\label{eq:v}
    \mathcal{V}(v) = \underbrace{\frac{1}{T} \int_0^T x^2(t)\,\mathrm{d}t}_{\mathcal{V}_1(v)} 
    - 
    \underbrace{\left( \frac{1}{T} \int_0^T x(t) \,\mathrm{d}t \right)^2}_{\mathcal{V}_2(v)}, 
\end{equation}
where $\mathcal{V}_1(v)$ and $\mathcal{V}_2(v)$ denote the mean square value and square of mean of the trajectory $x(t)$ over the duration of $[0,T]$, respectively. First, substituting the kinematic relation $x(t) = \int_0^t v(u) \,\mathrm{d}u$ into $\mathcal{V}_1(v)$ gives
\begin{align}
    \mathcal{V}_1(v) &= \frac{1}{T} \int_0^T \left( \int_0^t v(u) \, \mathrm{d}u \right)^2 \,\mathrm{d}t \nonumber \\
      &= \frac{1}{T} \int_0^T \int_0^t \int_0^t v(u) v(s) \,\mathrm{d}u \,\mathrm{d}s \,\mathrm{d}t,
\end{align}
where the region of integration is defined as $\left\{(u,s,t)\,\vert\, 0 \le u \le t, 0 \le s \le t, 0 \le t \le T \right\}$. Interchanging the order of integration, the region is equivalent to $\left\{(u,s,t) \,\vert\, u, s \in [0,T], t \ge \max(u,s) \right\}$. Thus, $\mathcal{V}_1(v)$ can be further derived as
\begin{equation}
\begin{aligned}
    \label{eq:v1}
    \mathcal{V}_1(v) &= \frac{1}{T} \int_0^T \int_0^T v(u) v(s) \left( \int_{\max(u,s)}^T \,\mathrm{d}t \right) \mathrm{d}u \, \mathrm{d}s  \\
      &= \int_0^T \int_0^T v(u) \frac{T - \max(u,s)}{T} v(s)  \, \mathrm{d}u \, \mathrm{d}s.
\end{aligned}
\end{equation}

Similarly, applying the same interchanging technique, the square of mean term $\mathcal{V}_2(v)$ can be rewritten as
\begin{equation}
\begin{aligned}
\label{eq:v2}
    \mathcal{V}_2(v) &= \frac{1}{T^2} \left( \int_0^T v(u) \left( \int_{u}^T \,\mathrm{d}t \right) \,\mathrm{d}u \right)^2 \\
      &= \int_0^T \int_0^T v(u) \frac{(T-u)(T-s)}{T^2} v(s)  \, \mathrm{d}u \, \mathrm{d}s.
\end{aligned}
\end{equation}

Therefore, substituting \eqref{eq:v1} and \eqref{eq:v2} into \eqref{eq:v}, the variance functional admits a compact kernel representation as
\begin{equation}
\label{eq: kernel_rep_v1}
    \mathcal{V}(v) = \iint\limits_{[0,T]^2} v(u) K(u,s) v(s)  \, \mathrm{d}u \, \mathrm{d}s,
\end{equation}
where the symmetric kernel function $K(u,s)$ is defined as
\begin{equation}
    K(u,s) = \frac{u+s - \max(u,s)}{T} - \frac{us}{T^2}.
\end{equation}

Noting that $u+s - \max(u,s) = \min(u,s)$, we obtain the expression in \eqref{eq:kernel} which completes the proof.

\section{Derivation of Jacobian Matrix}
\label{sec:app3}
To derive the $(k,j)$-th entry of $\mathbf{J}(\mathbf{c})$, we need to consider the $k$-th component $F_k(\mathbf{c})$ derived in \eqref{eq:nonlinear_function} and differentiate it with respect to the $j$-th coefficient $c_j$, one has
\begin{equation}
\label{eq:jacobian_derivation}
\begin{aligned}
\frac{\partial F_k}{\partial c_j} &= \frac{\partial}{\partial c_j} \left( 2(\lambda_k - \xi \alpha_1) c_k - 3\xi \alpha_2 \sum_{n=1}^N \sum_{m=1}^N c_n c_m \mathcal{T}_{nmk} \right) \\
&= A_{kj}\delta_{kj} - 3\xi \alpha_2 \sum_{n=1}^N \sum_{m=1}^N \frac{\partial (c_n c_m)}{\partial c_j} \mathcal{T}_{nmk},
\end{aligned}
\end{equation}
where $A_{kj}$ denotes the $(k,j)$-th entry of the diagonal matrix $\mathbf{A}$. Applying the product rule, the derivative of the quadratic term becomes
\begin{equation}
\begin{aligned}
\label{eq: grad_drag}
 \sum_{n=1}^N \sum_{m=1}^N ( \delta_{nj} c_m + c_n \delta_{mj} ) \mathcal{T}_{nmk} &= \sum_{m=1}^N c_m \mathcal{T}_{jmk} + \sum_{n=1}^N c_n \mathcal{T}_{njk} \\
&= \sum_{n=1}^N c_n (\mathcal{T}_{jnk} + \mathcal{T}_{njk}).
\end{aligned}
\end{equation}

Exploiting the symmetry of $\mathcal{T}$, i.e., $\mathcal{T}_{jnk} = \mathcal{T}_{njk}$, and substituting \eqref{eq: grad_drag} into \eqref{eq:jacobian_derivation}, the expression can be simplified as
\begin{equation}
\frac{\partial F_k}{\partial c_j} = A_{kj}\delta_{kj} - 6\xi \alpha_2 \sum_{n=1}^N c_n \mathcal{T}_{njk}.
\end{equation}

Consequently, the complete Jacobian matrix can be constructed analytically as
\begin{equation}
\mathbf{J}(\mathbf{c}) = \mathbf{A} - 6\xi \alpha_2 \mathbf{K}(\mathbf{c}),
\end{equation}
which completes the derivation.

\end{appendices}

\bibliographystyle{IEEEtran}
\bibliography{IEEEabrv,references}

\end{document}